\documentclass[dvips]{acta}

\usepackage{supertabular,lscape,epsfig}
\usepackage{amssymb}
\usepackage{amsmath}
\SetPages{1}{7}
\SetVol{62}{2012}

\begin{document}
\begin{Titlepage}

\Title { On the Periods of Negative Superhumps 
             \\ and the Nature of Superhumps }

\Author {J.~~S m a k}
{N. Copernicus Astronomical Center, Polish Academy of Sciences,\\
Bartycka 18, 00-716 Warsaw, Poland\\
e-mail: jis@camk.edu.pl }

\Received{  }

\end{Titlepage}

\Abstract { Osaki and Kato (2012) interpreted variations of the negative 
superhump periods, discovered by them in dwarf nova V1504 Cyg, as evidence 
in favor of the thermal-tidal instability model for superoutbursts. 
It is shown that their interpretation was incorrect. The observational 
evidence is recalled showing that superoutbursts are due to the enhanced 
mass transfer rate.  
} 
{accretion, binaries: cataclysmic variables, stars: dwarf novae }

\section { Introduction } 

In their recent paper Osaki and Kato (2012) presented results of the 
analysis of the Kepler light curve of the dwarf nova V 1504 Cygni.  
In particular they detected negative superhumps (nSH) and variations 
of their period. Interpreting those variations as being {\it exclusively} 
due to the variable radius of the disk they concluded that they provide 
support for the thermal-tidal model of superoutbursts. 

The purpose of the present paper is to challenge those conclusions. 
In Section 2 it will be shown that their interpretation of the nSH period 
variations and -- consequently -- their main conclusion about the nature 
of superoutbursts were incorrect. In Section 3 the observational evidence 
will be recalled showing that superoutbursts are due to the enhanced mass 
transfer rate.

\section { The Negative Superhump Periods and their Variations } 

\subsection {Negative superhumps in V1504 Cyg}

Osaki and Kato (2012) discovered negative superhumps (nSH) in the Kepler light 
curve of V1504 Cyg and studied them in detail during supercycle No.5. 
Results, presented in their Fig.5, show that the nSH frequency $\nu_{nSH}$ 
increased during outbursts and superoutbursts and decreased during 
quiescence. In addition, the mean frequency, averaged over 
normal outburst cycles, increased during the supercycle. 

According to the commonly adopted interpretation, negative superhumps 
are produced by the stream impact, as it transits across the face of the 
tilted, precessing disk. Adopting this interpretation Osaki and Kato assumed 
that the observed variations of the nSH frequency (or period) are 
{\it exclusively} due to the variations of the disk radius. 
In particular, they translated variations in $\nu_{nSH}$ into variations 
in $r_d$ using the following relation (from Larwood 1998) 

\beq
{{\nu_{nSH}}\over {\nu_{orb}}}~=~{{P_{orb}}\over {P_{nSH}}}~=~
1~+~{3\over 7}~ {q\over {(1+q)^{1/2}}}~ \cos \theta~ r_d^{3/2}~,  
\eeq

\noindent
where $\theta$ is the disk tilt angle and $r_d=R_d/A$. 

Variations of the nSH frequency in V1504 Cyg, when interpreted as being 
due to the variations of disk radius, are strikingly similar to the disk 
radius variations predicted by the thermal-tidal instability (TTI) model 
(Osaki 1989,2005). 
It is therefore not surprising that Osaki and Kato interpreted this 
similarity in favor of the TTI model. 

Upon closer examination, however, it turns out that their interpretation 
of the nSH period variations leads to several discrepancies and inconsistencies 
and therefore cannot be correct. Consequently their main conclusion 
about the nature of superoutbursts is also incorrect.

\subsection {Amplitudes of negative superhumps } 

Under the assumption that the mass transfer rate is constant, the nSH 
amplitude depends only on the amount of the kinetic energy of the stream 
dissipated during its collision with the disk, which is 
proportional to the (dimensionless) impact parameter $\Delta$v$^2$. 
From calculations of the stream particle trajectories one finds that 
the value of this parameter increases with decreasing radial distance. 
This implies that the nSH amplitude must be {\it smaller} when the disk is 
{\it larger} and {\it vice versa}. 

Osaki and Kato (2012, Figs.7 and 8) determined the average nSH light curves 
and amplitudes in two cases: 

{\parskip=0truept {
(1) During the superoutburst No.4 the amplitude was $A=0.04$ mag, or 
-- corrected for the $\sim 3$ mag difference in brightness between 
superoutburst maximum and quiescence -- $A_{corr}\approx 0.6$ mag. 
The nSH period, used by Osaki and Kato to calculate the average light curve, 
was $P_{nSH}=0.067764$d, which corresponds -- {\it via} Eq.(1) -- to 
$r_d=0.48$. 

(2) During the 8-day interval at quiescence (BJD 2455440-448) the amplitude 
was $A=0.35$ mag. The nSH period used by Osaki and Kato was in this case 
$P_{nSH}=0.068076$d implying $r_d=0.42$. 

Obviously then in case (2), when the disk was smaller, the nSH amplitude 
should be {\it larger} than in case (1). Instead it was much {\it smaller}. 
}}
\parskip=12truept 

\subsection {Negative superhumps during normal outburst cycles } 

Fig.5 in Osaki and Kato (2012) shows that during all normal outburst cycles 
the minima of $\nu_{nSH}$ occured $\sim 3$ days {\it before} the initial 
rise to outburst maximum and the following increase of $\nu_{nSH}$ till 
its maximum lasted for $\sim 3$ days. 
Within their interpretation this would imply that the disk begins 
to expand $\sim 3$ days {\it before} the initial rise and continues to expand 
for about 3 days. 

Model calculations for dwarf nova outbursts show, however, that the expansion 
of the disk occurs nearly {\it simultaneously} with rising light and -- in 
the case of short period systems -- lasts for only $\sim 0.5$ day (see, for 
example, Fig.3 in Hameury et al. 1998, or Fig.1 in Osaki and Kato 2012). 
The interpretation is simple: Accretion of the material from the outer parts  
of the disk causes its luminosity to increase, while the excess angular 
momentum transported outward causes the disk to expand; this occurs on 
a viscous time scale which -- in this case -- is shorter than 1 day. 
Obviously then the expansion of the disk cannot begin 3 days earlier 
and cannot last 3 days.

\subsection {Negative superhumps during superoutbursts } 

Fig.5 in Osaki and Kato (2012) shows that during the main part of both 
superoutbursts, lasting for about 8 days, the nSH frequency decreased 
from $\nu_{nSH}=14.78$ to $\nu_{nSH}=14.70$. Using Eq.(1) we get 
for the corresponding disk radii: $r_d=0.51$ and $r_d=0.44$ and, 
consequently, $dr_d/dt=-0.007$ d$^{-1}$. 

For comparison we have the radii of the disk in dwarf nova Z Cha during 
its superoutbursts determined from eclipses of the hot spot (Appendix). 
The least squares fit to those data gives 

\beq
r_{d,\circ}~=~0.481\pm 0.016,~~~~~{\rm and}~~~~~
      {{dr_d}\over {dt}}~=~0.0003\pm 0.0043~{\rm d}^{-1}~, 
\eeq

\noindent 
with no difference between $r_d(\phi_i)$ and $r_d(\phi_e)$. 
This shows that the radius of the disk remains {\it constant} throughout   
superoutburst. Worth noting is also that $r_d\approx r_{tidal}$, 
as expected in the case of steady-state accretion with high mass 
transfer/accretion rate.

\subsection {Comparison with other systems }

Negative superhumps and their variable periods have been observed in several 
other dwarf novae at various stages of their outburst and superoutburst cycles. 
The best documented examples are: 

{\parskip=0truept {
{\it V503 Cyg}. Harvey et al. (1995) detected negative superhumps during 
quiescence as well as during outbursts and superoutbursts. 
According to them the nSH period was longer during quiescence and 
shorter during outbursts, which is similar to the situation observed 
in V1504 Cyg. However, unlike in the case of V1504 Lyr, the nSH period, 
averaged over normal outburst cycles, did not show any significant change 
between the two successive superoutbursts. 

{\it BK Lyn}. The light curve and the (O-C) diagram in Kato et al. 
(2012, Fig.26) show that between the two successive superoutbursts 
the nSH period decreased -- like in V1504 Cyg. On the other hand, 
however, unlike in the case of V1504 Cyg, no variations were present 
during normal outburst cycles. 

{\it V344 Lyr}. Wood et al. (2011) detected negative superhumps in the Kepler 
light curve of this star, present occasionally during some of the quiescence 
intervals and during some of normal outbursts. 
As can be seen from their Figs.23 and 26, the behavior of the nSH period was 
similar to that of V1504 Cyg: it was increasing during quiescence, 
decreasing during normal outbursts, and decreasing during supercycles. 

{\it ER UMa}. The light curve and the (O-C) diagram in Ohshima et al. 
(2012, Fig.2) show that the nSH period variations in this system are similar 
to those in BK Lyn. 
}}
\parskip=12truept 

Those examples show that (1) the decreasing nSH period during supercycles 
is a common phenomenon among dwarf novae with superoutbursts, while 
(2) the nSH period variations during their normal outbursts cycles 
occur only in some of them. In particular, the nSH period variations 
observed in V1504 Cyg cannot be considered as representative for 
all such systems.

\subsection {Negative superhumps periods and their variations}

In the simplest case, when both -- the precession period and the 
nSH period are constant, we have  

\beq
{1 \over {P_{nSH,\circ}}}~=~{1 \over {P_{prec}}}~+~{1 \over {P_{orb}}}~.    
\eeq

The precession period depends not only on the effective radius of the 
disk but also on the distribution of its surface density $\Sigma (r)$
(cf. Larwood 1998, Montgomery 2009 and references therein) 

\beq
P_{prec}~=~f\left[r_d,\Sigma(r)\right]~, 
\eeq 

\noindent
where both -- $r_d$ and $\Sigma(r)$ are known to change during the dwarf 
nova cycle. 

Turning to the {\it observed} nSH period we must note that it is defined 
as the interval of time between two successive maxima resulting from the
collision of the stream with the surface of the tilted disk. 
The effective location of the stream impact depends on the disk tilt 
and on its geometrical thickness, described by $z/r=f(r)$, which also  
changes during the dwarf nova cycle. Taking this into account we can write 

\beq
P_{nSH,obs}~=~P_{nSH,o}~\left ( 1~+~{{d\Delta t}\over {dt}} \right )~,  
\eeq

\noindent
where 

\beq
\Delta t~=~f\left[\theta,z/r(r)\right]~ 
\eeq 

\noindent
is the "flight" time of the stream elements from L$_1$ to the effective 
point of collision. 

Combining Eqs.(3-6) we obtain the following general formula describing 
the nSH period variations

\bdm
{{dP_{nSH,obs}}\over {dt}}~=~
\epsilon_{nSH}~ {{\partial P_{prec}}\over {\partial r_d}}~{{dr_d}\over{dt}}~+~
\epsilon_{nSH}~ {{\partial P_{prec}}\over {\partial \Sigma (r)}}~
{{d\Sigma (r)}\over{dt}}~
\edm

\vskip -5truemm

\beq
+~\epsilon_{nSH}~ {{\partial P_{prec}}\over {\partial r_d}}~{{dr_d}\over{dt}}~
{{d\Delta t}\over {dt}}~+~
\epsilon_{nSH}~ {{\partial P_{prec}}\over {\partial \Sigma (r)}}~
{{d\Sigma (r)}\over{dt}}~{{d\Delta t}\over {dt}}~
+~P_{nSH,\circ}~{{d^2\Delta t}\over {dt^2}}~,  
\eeq

\noindent
where $\epsilon_{nSH}=1-P_{nSH}/P_{orb}$. 

This shows that the nSH period variations can be due to many different causes.
Osaki and Kato limited their discussion only to the first term on the 
right-hand side of Eq.(7) thereby making their conclusions unreliable.

\section { Evidence for the Enhanced Mass Transfer Rate } 

The presence and luminosity of the hot spot in Z Cha (Smak 2007,2008a) 
and in other dwarf novae (e.g. OY Car; Smak 2008b) during their 
superoutbursts provide direct observational evidence showing that 
superoubursts are due to strongly enhanced mass transfer (EMT) rate. 

Osaki and Kato ignore this evidence. Instead they write: 
{\it "In the EMT model mass-transfer rate {\underbar {is thought}} to be 
greatly increased during a superoutburst"}. 
Furthermore, in their Table 1, the {\it "enhanced hot spot"} is listed   
as a {\it "consequence"} of the model, which -- according to them -- 
is {\it "not in agreement with observation"}!

\vskip 0.5cm

\centerline {\bf Appendix }

The radii of the disk in dwarf nova Z Cha during its superoutbursts were 
determined (Smak 2007) from phases of ingress and egress of the hot spot 
eclipses. The most reliable values of $r_d(\phi_i)$ and $r_d(\phi_e)$, 
obtained from eclipses observed at beat phases $\phi_b=0.4-0.6$ 
(which have not been published before) are listed in Table 1. 
The parameter $\delta t$ is the time (in days) since the beginning of 
superoutburst. 

\begin{table}[h!]
{\parskip=0truept
\baselineskip=0pt {
\medskip
\centerline{Table 1}
\medskip
\centerline{ Disk radii in Z Cha during superoutbursts }
\medskip
$$\offinterlineskip \tabskip=0pt
\vbox {\halign {\strut
\vrule width 0.5truemm #&	
\enskip\hfil#\hfil\enskip&      
\vrule#&			
\enskip\hfil#\hfil\enskip&      
\vrule#&			
\enskip\hfil#\hfil\enskip&      
\vrule#&			
\hfil#\hfil&	                
\vrule#&			
\hfil#\hfil&	                
\vrule width 0.5 truemm # \cr	
\noalign {\hrule height 0.5truemm}
&&&&&&&&&&\cr
&Eclipse&&$\delta t$&&$\phi_b$&&$r_d(\phi_i)$&&$r_d(\phi_e)$&\cr
&&&&&&&&&&\cr
\noalign {\hrule height 0.5truemm}
&&&&&&&&&&\cr
& E54023 && 2.5 && 0.56 && 0.48 && 0.39 &\cr
& E54024 && 2.5 && 0.56 && 0.46 && 0.43 &\cr
& E54025 && 2.5 && 0.56 && 0.46 && 0.54 &\cr
& E54076 && 6.5 && 0.41 && 0.45 && 0.50 &\cr
& E54077 && 6.5 && 0.41 && 0.46 && 0.51 &\cr
& E59875 && 4.5 && 0.46 && 0.50 && 0.47 &\cr
& E67837 && 1.5 && 0.55 && 0.52 && 0.51 &\cr
& E67838 && 1.5 && 0.55 && 0.46 && 0.49 &\cr
& E67864 && 3.5 && 0.45 && 0.51 && 0.48 &\cr
& E77865 && 1.5 && 0.55 && 0.51 && 0.47 &\cr
&&&&&&&&&&\cr
\noalign {\hrule height 0.5truemm}
}}$$
}}
\end{table}

\begin {references} 

\refitem {Hameury, J.-M., Menou, K., Dubus, G., Lasota, J.-P., Hur{\'e}, J.-M.} 
      {1998} {\MNRAS} {298} {1048} 

\refitem {Harvey, D., Skillman, D.R., Patterson, J., Ringwald, F.A.} 
      {1995} {\PASP} {107} {551} 

\refitem {Kato, T. {\it et al.}} {2012} {\rm arXiv: 1210.0678} {~} {~} 
 
\refitem {Larwood, J.} {1998} {\MNRAS} {299} {L32} 

\refitem {Montgomery, M.M.} {2009} {\ApJ} {705} {603}

\refitem {Ohshima, T. {\it et al.}} {2012} {\it Publ.Astr.Soc.Japan} {64} {L3} 

\refitem {Osaki, Y.} {1989} {\it Publ.Astr.Soc.Japan} {41} {1005}

\refitem {Osaki, Y.} {2005} {\it Proc.Japan Academy, Series B} {81} {291}

\refitem {Osaki, Y., Kato, T.} {2012} {\rm arXiv: 1212.1516} {~} {~} 

\refitem {Smak,J.} {2007} {\Acta} {57} {87}

\refitem {Smak,J.} {2008a} {\Acta} {58} {55}

\refitem {Smak,J.} {2008b} {\Acta} {58} {65}

\refitem {Wood, M.A., Still, M.D., Howell, S.B., Cannizzo, J.K., Smale, A.P.}
      {2011} {\ApJ} {741} {105}

\end {references}

\end{document}